\documentclass[prd, twocolumn, nofootinbib, floatfix]{revtex4-2} 

\usepackage{amsmath}
\usepackage{graphicx}
\usepackage{dcolumn}
\usepackage{bm}
\usepackage{epsfig}
\usepackage{amssymb,latexsym,mathrsfs,empheq}
\usepackage{graphicx}
\usepackage{color}
\usepackage{xcolor} 
\usepackage{mathtools} 
\usepackage{hyperref} 
\usepackage{cleveref}

\hypersetup{
    colorlinks=true,
    linkcolor=red,
    citecolor=blue,
} 

\newcommand{\be}{\begin{equation}}
\newcommand{\ee}{\end{equation}}
\newcommand{\bs}{\begin{split}} 
\newcommand{\bea}{\begin{eqnarray}}
\newcommand{\eea}{\end{eqnarray}}

\begin{document}

\title{Interpreting Dark Energy Data Away from $\Lambda$} 

\author{Eric V.\ Linder} 
\affiliation{
Berkeley Center for Cosmological Physics \& Berkeley Lab, 
University of California, Berkeley, CA 94720, USA
} 

\begin{abstract} 
Dark energy away from a cosmological constant $\Lambda$ -- like 
early universe inflation that ends -- can be understood in terms 
of well defined physical behaviors. These guide dark energy into 
thawing or freezing classes, with $w_0$--$w_a$ arising as a 
physical calibration of the phase space. Other regions of phase space 
-- zones of avoidance -- require violation of some basic principle. 
We explore these cases, drawing a direct analogy with how nonGaussianity 
in inflation can add physics beyond standard dynamics. 
We examine the physics implications if the 
best fit of current data is taken to be truth, outlining four properties, and investigate 
the reality 
of phantom crossing $w=-1$, finding it significantly favored.  
\end{abstract} 

\date{\today} 

\maketitle

\section{Introduction} 

Cosmic acceleration is amazing. It bends the cosmic expansion 
against the attraction of gravity, stretching distances, and 
suppresses growth of large scale structure, again against the 
attraction of gravity, possibly shutting growth off in the future. We 
do not know its source other than that it lies outside the 
Standard Model of particle physics. 

The ``simplest'' explanation might be a cosmological constant $\Lambda$
but this comes with its own mysteries and paradoxes, such as why 
it has neither Planck scale energy density nor  zero but rather of such 
a magnitude that it only became significant within the last efold 
of expansion history -- moreover despite quantum corrections that 
should enter at higher energy scales. 

Going beyond $\Lambda$ gives dark energy some dynamics, and this 
carries with it clear physical characteristics due to the dark 
energy field existing in a universe with many efolds of radiation and matter 
dominated expansion. Thus it is very comprehensible. (For expansion 
behavior, even an origin such as a modification of gravity can be 
treated adequately as an effective scalar field with potential and 
kinetic terms.) 

In \Cref{sec:dyn} we briefly review dark energy dynamics and 
its resulting physical characteristics. \Cref{sec:mirage} considers 
current data at face value and its implications of going beyond 
standard principles, with \Cref{sec:nonKG} illustrating how this 
tests the framework itself, in close analogy with inflation. 
We look to 
the future in \Cref{sec:fut} for what other types of data can 
say about the physical origin of dark energy.

\section{Physics of Dark Energy Dynamics} \label{sec:dyn} 

Dark energy does not exist in a vacuum. Less waggishly, we 
understand the basics of dark energy behavior very well because 
it has evolved not alone but in an environment of many efolds of 
radiation and matter dominated expansion. This forces it to have 
certain dynamics that cannot be avoided without fine tuning (or 
violating some standard principles, as discussed in the next 
section).  

The immense Hubble friction from the environment governs the 
dark energy dynamics, together with steepness of the dark energy 
potential. Put most simply (see \cite{caldlin,paths,flow} for 
details) if the Hubble friction dominates then the (effective) 
field is frozen in place, released to thaw only when the universe 
expands sufficiently to dilute radiation and matter and reduce 
the Hubble friction. Such thawing fields evolve away from cosmological 
constant behavior. If instead the potential has sufficiently steep slope, the field rolls in the early universe until eventually 
it approaches a region of shallow slope near the potential minimum. 
The Hubble friction takes over and the field slows. Such freezing 
fields evolve toward cosmological constant behavior. 

One could plot these behaviors in field phase space $\phi$--$\dot\phi$ 
but it is more convenient to work with the dark energy pressure to 
energy density, or equation of state, ratio $w$ and its efold derivative 
$w'$. This is closer to the Hubble parameter $H=\dot a/a$ describing 
cosmic expansion history since 
\be 
\left(\ln H^2\right)'\equiv\frac{d\ln H^2}{d\ln a}=
-3\left[1+\sum w_i(a)\,\Omega_i(a)\right]\ .  
\ee 
Furthermore, it is cleaner if the dark energy is not actually 
a scalar field. 

\begin{figure} 
\centering 
\includegraphics[width=\columnwidth]{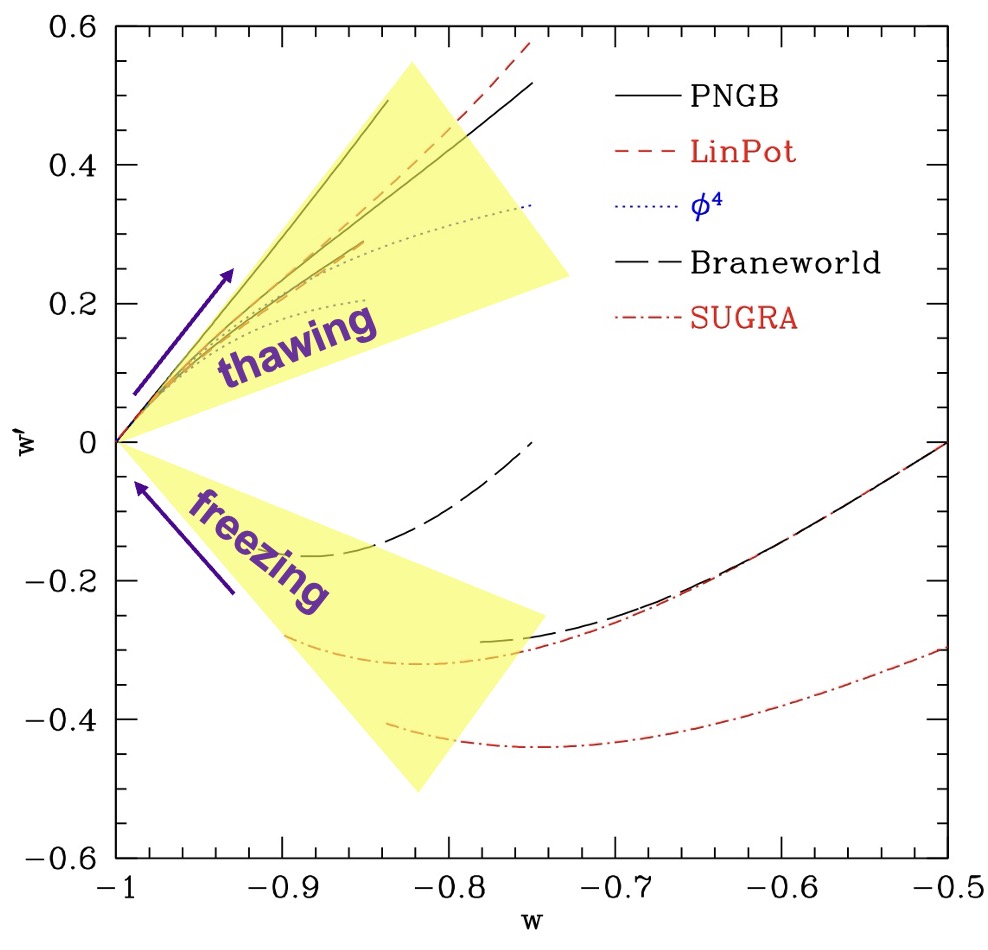}
\caption{Simple physics in an expanding universe brings 
dark energy dynamics into narrow regions of the phase 
space, either thawing or freezing. Adapted 
from \cite{calde}.}
\label{fig:wwp} 
\end{figure}

\Cref{fig:wwp} illustrates how these physical effects of Hubble friction 
and potential steepness, operating over many efolds of radiation and 
matter domination, bring the dark energy evolution into narrow regions 
of the phase space, i.e.\ the thawing and freezing classes.

\begin{figure*} 
\centering 
\includegraphics[width=\textwidth]{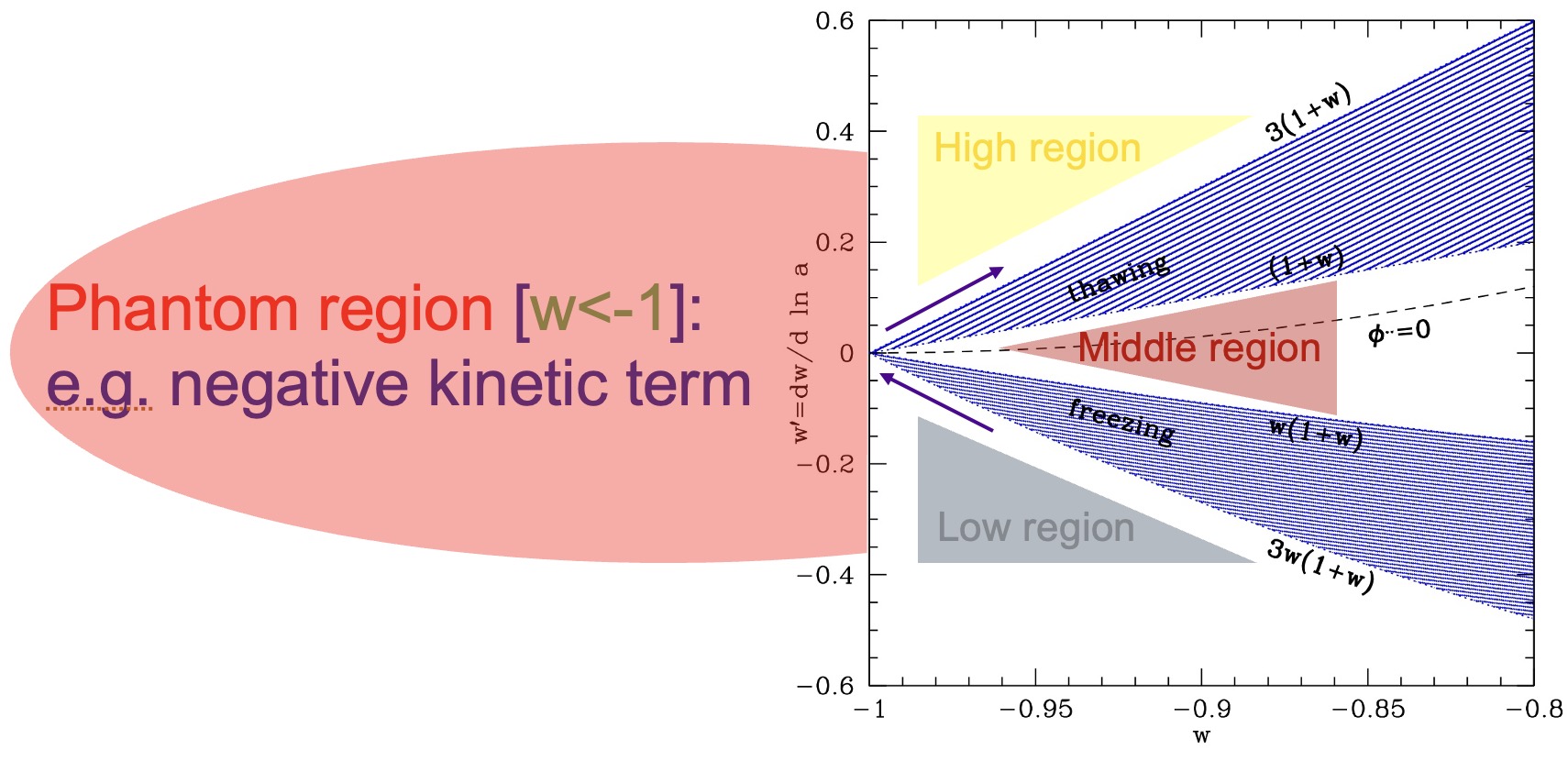}
\caption{Zones of avoidance, outside the thawing and 
freezing regions, can only be accessed by violating some 
basic physical principle. The text describes these for 
each region.}
\label{fig:zone} 
\end{figure*}

What about the rest of phase space? In order for dark energy at 
present to lie in another region, some standard physical principle 
must be broken. \Cref{fig:zone} defines the ``zones of avoidance''. 
The middle region, between the thawing and freezing regions, requires 
fine tuning, where the two physical effects must -- coincidentally 
just at present -- nearly balance. The high region violates early 
radiation and matter domination, i.e.\ the many efolds of Hubble 
friction. The low region requires the field to evolve noncanonically, 
e.g.\ roll upslope. On the left half of the diagram the field is in 
the phantom regime, $w<-1$, e.g.\ with a negative kinetic term. 
We will revisit the phantom and high regions in the next section.

Thus we know quite a lot about dark energy characteristics even 
without identifying dark energy in specific. The $w$--$w'$ phase 
space provides an excellent guide to the physics affecting dark 
energy. However, cosmological data cannot generally fit the two 
full functions $w(a)$, $w'(a)$. Indeed, observations depend on 
(multiple) integrals over $w(a)$ so the data does not see the details 
of the functions. Principal component analysis, or similar techniques, 
demonstrate that only two numbers, not full free functions, can be 
fit \cite{pca}, at least until observations improve beyond 0.1\% 
precision. 

\begin{figure*} 
\centering 
\includegraphics[width=\textwidth]{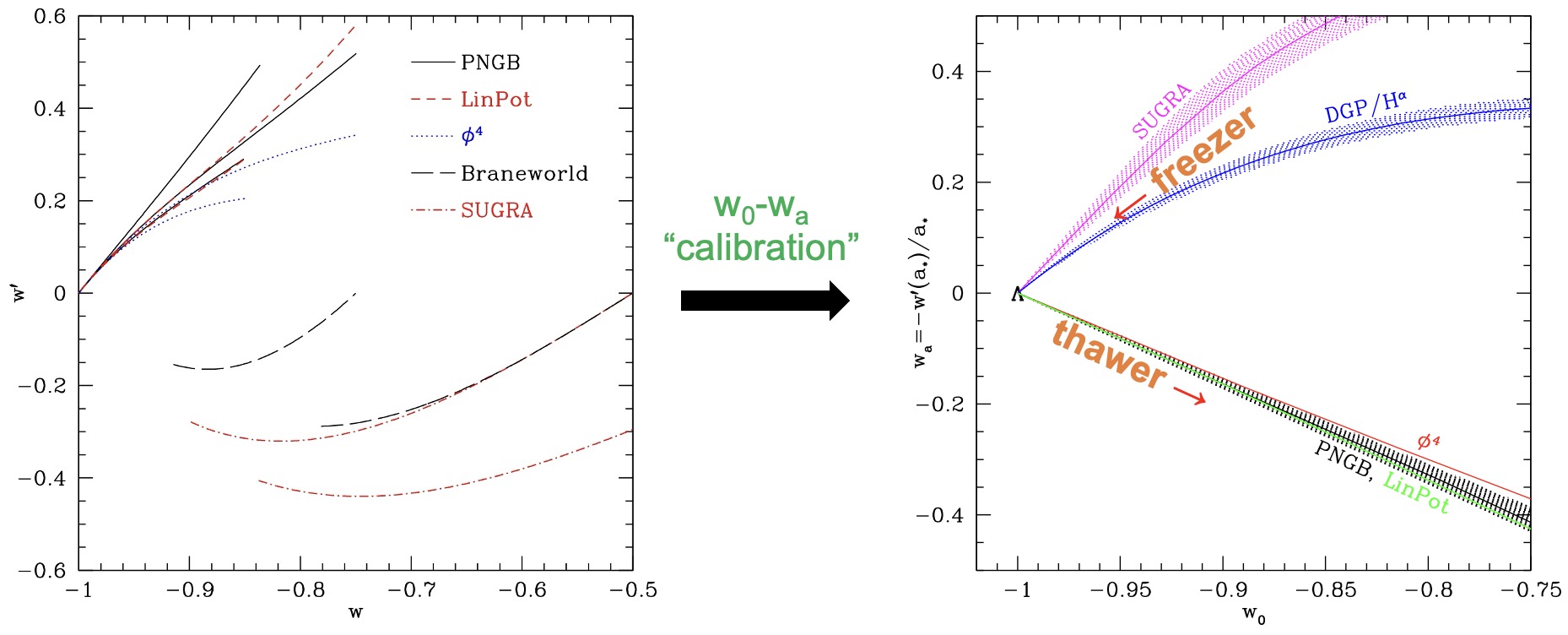}
\caption{Dark energy evolution can be further calibrated 
from the thawing/freezing wedges in $w$--$w'$ of 
\Cref{fig:wwp} [see left panel] into tight, near universal 
behavior through stretching the time coordinate [see right 
panel]. This physical calibration is the foundation of 
$w_0$--$w_a$. (Note the sign flip of $w_a$ 
relative to $w'$, interchanging the top and 
bottom half planes.) Adapted from \cite{calde}.}
\label{fig:wa} 
\end{figure*}

The art, then, is choosing the right two quantities to preserve  
the essential physics information. For our parameter corresponding 
to $w(a)$, let us choose the value today, $w_0\equiv w(a=1)$. To 
describe $w'(a)$, let's try stretching the time axis, i.e.\ scaling 
$w'(a)$. \Cref{fig:wa} shows the results -- a remarkable success! 
Different varieties of theories are tightly calibrated, and in 
particular the wide variety of different models in the thawing class 
come together to form a universal behavior in this new phase space. 
(See \cite{calde} for more details.)

This is the meaning of $w_0$--$w_a$, with $w(a)=w_0+w_a(1-a)$. 
Created as a fit to exact 
solutions of scalar field dynamics in \cite{linexp} -- never as 
an arbitrary parametrization or ``Taylor expansion'' -- it then 
was demonstrated to be a calibration form, a physics based 
encapsulation of the full $w$--$w'$ phase space in \cite{calde}.

\section{Interpreting Dark Energy Data} \label{sec:mirage} 

Now that we have the tools in place: that $w_0$--$w_a$ faithfully 
describes dark energy physics, what standard principles predict in 
terms of the expected regions of thawing and freezing, and what 
violations must occur to avoid those regions, let us consider the 
current data. 

The Dark Energy Spectroscopic Instrument (DESI \cite{desi}) recently 
released cosmological analyses \cite{kp7,crossing,focus} of its 
highly precise first year baryon acoustic oscillation (BAO) distance 
measurements, supplemented with supernova (SN) distances and cosmic 
microwave background (CMB) data. The analyses -- starting with blinded 
BAO data -- were done in several different ways and all came to a 
similar conclusion: $\Lambda$CDM is disfavored at $\sim2.5$--3.9$\sigma$ 
and the best fit lies neither in the thawing nor freezing regions, but 
what we here call the high region. 

The core principle violated in the high region is that of many efolds 
of dark energy evolution in a radiation and matter dominated universe. 
This is quite astounding, as that is the principle we might consider 
least likely to be broken (as opposed to, say, the scalar field 
kinetic structure), since it is indicated by a vast array of 
cosmological observations. One possibility is that there are 
statistical fluctuations or systematic offsets in the data, but 
this is somewhat alleviated by the blind nature of the analysis 
and the robustness of the result under different analysis approaches. 

Therefore we will focus here on the implications if the 
dark energy properties are truly as indicated by the best fit values, 
say $w_0\approx-0.64$, $w_a\approx-1.27$. (We will not worry here 
about specific values, only about what region of the phase space. 
And we will keep in mind the ``if'': recall that $3\sigma$ is not 
definitive, and the DESI analysis mindfully used the phrasing ``tantalizing 
hint of deviations'' from $\Lambda$.) 

Taking the best fit as truth requires the dark energy to possess 
four properties: 
\begin{itemize} 
\item The dark energy was phantom at $z\gtrsim1$; 
\item The dark energy must ``superevolve'' faster than the Hubble 
friction from matter domination would indicate; 
\item The dark energy crossed $w=-1$, ``the phantom divide''; 
\item The dark energy recently evolved toward $w>-1$, i.e.\ away from 
$\Lambda$. 
\end{itemize} 
We know the physics to give dark energy each individual property, 
but they generally don't all go together! 

Beyond all these aspects concerning the general region of phase space, 
the best fit lies along a very special cut through phase space 
call mirage dark energy. Mirage dark energy is a continuous 
family of dark energy 
that extends from $\Lambda$ into the high region, along the 
line $w_a\approx -3.66(1+w_0)$ \cite{mirage}. It has several interesting 
properties, including that the distance to CMB last scattering is 
nearly identical for all members of the family (and so if one 
naively fit only a constant $w$, rather than looking at the 
dynamics, one would find $\langle w\rangle\approx-1$). 

In seeking the physics behind the dark energy, an important 
question is whether the apparent interpretation regarding 
the phantom crossing is robust. That is, does $w(a)$ really 
cross $-1$ or does it just appear that way because of the values 
of $w_0$--$w_a$? This has been partly addressed in 
\cite{crossing} by using a nonparametric $w(a)$, and in 
\cite{focus} by using a $w(a)$ that does not cross $-1$, showing that the crossing is real, 
but here we revisit it more quantitatively. 

We know that values of $w_0$--$w_a$ characteristic of the 
thawing region (not the high region) can be treated by the 
algebraic thawing form \cite{algthaw} with $w(a)>-1$ always, with 
the two approaches agreeing to ${\mathcal O}(0.1\%)$. 
However, this does not 
work for the strong mirage case where $|w_a|\gtrsim 1$. 
Using the algebraic thawing form with $p=5$ (rather than the 
$p=1$ for thawing dark energy, in order to give a sharp 
evolution and hence large $|w_a|$) yields a $w(a)>-1$ at all 
times and yet $w_0$--$w_a$ parameters that approximate our 
best fit values. (See \cite{steinh} for a similar approach.) 

Fitting all the data to either the thawing with $w_0$--$w_a$  
or the algebraic thawing forms gives no discernible 
difference. Their $\Delta\chi^2<0.3$ (also see 
\cite{focus}). However, fitting all 
the data to the mirage or the algebraic form with 
$p=5$ yields a clear distinction. While the mirage case 
gives a superb fit (since the best fit to the data indeed 
lies nearly on the mirage line), the algebraic form -- 
where phantom crossing is merely a spurious interpretation 
due to $w_0$--$w_a$ -- suffers by $\Delta\chi^2>9$, 
i.e.\ $\sim3\sigma$ worse. Thus the data favors that 
the phantom crossing is indeed a real property.

\section{Interpreting Dark Energy Physics} \label{sec:nonKG} 

Thus the four conditions itemized above describe 
the physics needed. Before pursuing the specific physics 
indicated, let us look at the bigger picture. What we are 
really doing is testing the framework, not fitting 
parameters or models. Any cosmic expansion history can be 
treated as an effective scalar field, i.e.\ $w$--$w'$. So 
data giving a result in one of the Zones of Avoidance is 
testing the physics of the scalar field, not just one 
particular model. In gist, we are checking the dynamical 
(Klein-Gordon) equation itself. 

Note that this is very similar to what is done with early 
universe inflation. Beyond the dynamics (e.g.\ the scalar 
spectral tilt $n_s$ and primordial gravitational wave or 
tensor to scalar ratio $r$), we look at for example 
inflationary nonGaussianity. Here we are looking at dark 
energy nonKlein-Gordonness. 

As in inflation, this can arise through any of the 
following physical characteristics: 
\begin{itemize} 
\item Noncanonical kinetic term 
\item Multiple, interacting fields 
\item Nonstandard vacuum 
\end{itemize} 
A noncanonical kinetic term seems unlikely to be the 
(sole) cause of data preferring the high region in 
phase space, because such k-essence lives in the low 
region (or its mirror in the left half plane). Multiple, 
interacting fields is a possibility in fact richer than 
its inflation analog, since dark energy could interact 
not only with another scalar field but matter fields 
(which are negligible in the inflationary phase). 
While for inflation 
a nonstandard vacuum means not having an initial 
Bunch-Davies state, for dark energy it refers to the 
scalar field either having a negative potential or 
arising through a phase transition. 

To achieve all four physical conditions needed from 
the previous section, we can now see a possibility:  
dark energy that originates in a phase transition 
(e.g.\ vacuum metamorphosis \cite{vm1,vm2,vm3}) at 
redshift $z\approx1-2$ 
(thus avoiding the ``many efolds of dark energy evolution in a radiation and
matter dominated universe'' -- not because 
of violating matter domination but due to 
hiding from the many efolds), 
so it superevolves, rapidly 
increasing its energy density from near zero (so 
$w_a\lesssim-1$) and heading from deeply phantom 
toward $w=-1$. But we need something more: most 
phase transition models asymptote to a de Sitter 
future, with $w=-1$ (in the vacuum metamorphosis 
case this is because the Ricci scalar goes to a 
constant, locking into the value given by the 
scalaron mass). And indeed such ``emergent'' dark 
energy models do not fit the data as well as  
mirage dark energy (which crosses $w=-1$), e.g.\ 
\cite{focus}. 

The second element needed is some coupling that 
brings the field across $w=-1$ so that today 
$w_0>-1$. Indeed, particle physics generally gives 
interactions unless some symmetry or other principle 
suppresses it. Thus one possible physics 
framework for the 
desired mirage dark energy is a phase transition 
with the resulting field having a natural coupling. 
Such a theory would promise a rich array of tests 
from observational data.

\section{More Data for More Physics} \label {sec:fut} 

More data will not only test whether the best fit 
maintains its position in the high region of phase 
space, but can offer different angles on the physics. 
Just as in inflation one seeks to measure primordial 
gravitational waves and look for consistency between 
$r$ and $n_s$, for dark energy one can contrast growth 
probes with distance probes. 

More generally one can explore the gravitational sector, 
in terms of 
\begin{itemize} 
\item Growth vs expansion 
\item Gravitational effects on matter and light 
\item Gravitational friction on gravitational wave 
propagation 
\end{itemize} 
The second and third investigate gravity beyond general 
relativity, with the second looking for time and space 
dependence in the strength of gravitational coupling, 
i.e.\ $G_N\to G_{\rm matter}(k,a)$, $G_{\rm light}(k,a)$, 
and the third checking for a time variation of the 
Planck mass through a gravity friction term added to 
the Hubble expansion friction in gravitational wave 
propagation, causing the wave amplitude to decay 
differently than the standard 1/distance. 

Thus, measurements of growth of large scale structure, 
gravitational lensing, and gravitational wave distances 
from standard sirens all enable new tests of the 
physics framework. Note that the best fit we have been 
considering does not require any modification to general 
relativity though, and there may be no signal in the 
second and third items. 

General relativity does give a fixed relation between 
the growth factor of large scale structure and the 
cosmic expansion history. So growth observations such 
as from redshift space distortions and peculiar velocities 
can provide a test of the predicted consistency, whether 
the cosmology is that of the best fit, or of $\Lambda$. 
That is important, but if the growth data shows consistency 
with the best fit it does not necessarily zero in on 
the physical origin. 

Growth data does tend to constrain dark energy with a 
different degeneracy direction, i.e.\ weighting of $w_0$ 
vs $w_a$ say, than distance data. One can partly view 
this as that distances are measured from the observer 
back in history to the source, while growth depends on 
evolution from the early universe to the tracer. 

\begin{figure} 
\centering 
\includegraphics[width=\columnwidth]{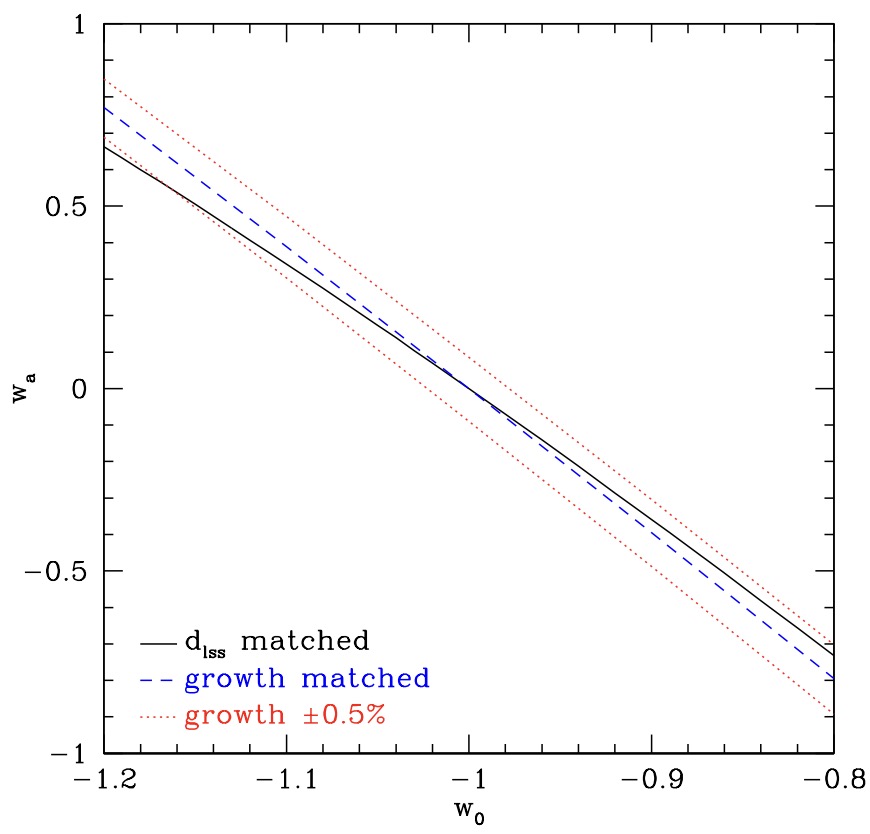}
\caption{Matching the CMB distance to last scattering is 
closely related to matching the growth history, in 
particular the linear growth factor ratio 
$g(a = 0.35)/g(a = 1)$. Thus for a universe involving 
mirage dark energy behavior, growth data such as from 
redshift space distortions would add only modest leverage 
to distance data. Adapted from \cite{mirage}.}
\label{fig:grocmb} 
\end{figure}

Thus the constraints on dark energy properties are 
often tighter when combining growth and distance data. 
However, there is one notable exception! Mirage dark 
energy -- exactly what the best fit seems to indicate -- 
is mostly exempt from this. Its particular properties 
give that the growth leverage direction essentially 
overlaps with that of the distance to CMB last scattering, 
and no major gain is expected. This is illustrated in 
\Cref{fig:grocmb}. So sadly, growth data will not have as 
much of an impact on cosmological constraints as it 
would in a universe that happened to be off the 
mirage line. 

Nevertheless, new data -- of expansion, growth, and 
gravity -- will be highly useful for crosschecking 
the current ``tantalizing hints'', testing general 
relativity, and gradually narrowing in on parameter 
values.

\section{Conclusions} 

Dark energy characteristics can be well understood as the 
result of physics during evolution through a long period 
of radiation and matter domination. The impact of dark 
energy on observations is superbly accounted for by 
means of the physical calibration of the phase space via 
$w_0$--$w_a$. 

We also understand the ``zones of avoidance'' in dark 
energy phase space and what standard principles would need 
to be violated to allow those regions to come into play. 

Amazingly, current data gives tantalizing hints that indeed 
dark energy behaves as if it lies in the high region zone 
of avoidance. Dark energy is best fit for current data by mirage dark energy, 
which has several unusual properties including phantom 
crossing and a non-de Sitter future state, and moreover 
exhibits a rapid evolution violating having a long history of 
evolution during matter domination. A possible physical 
scenario is identified of a phase transition, with the 
resulting field coupled in some manner to another (possibly  
matter) field. We emphasize that the same physical 
characteristics have been found in approaches not utilizing 
$w_0$--$w_a$ but more model agnostic forms. Furthermore 
we can verify that the phantom crossing being a real, not 
merely apparent, characteristic is preferred by the data 
by $\gtrsim3\sigma$. 

In analogy to nonGaussianity in inflation we can identify 
physical origins for violating the standard Klein-Gordon 
evolution. Nonstandard vacuum in the form of 
a phase transition is a contender. Similarly we can mirror inflationary consistency 
tests to reveal more about the nature of dark energy, 
especially through gravity beyond general relativity. 
We also note that due to the unique properties of mirage 
dark energy, growth of structure observations will tend 
to give more incremental rather than orthogonal 
improvements. 

Nevertheless, the data that is forthcoming of new and 
higher accuracy expansion, growth, and gravity observations 
will test the tantalizing hints of strong mirage dark 
energy. Other methods searching for a phase transition 
could yield interesting contributions. In particular, 
the next generation of experiments aiming for a higher 
redshift range, e.g.\ DESI2 \cite{desi2}, SpecS5 
\cite{specs5}, could explore whether dark energy density 
was vanishingly small at $z\gtrsim2$ \cite{rise,sailer}, 
indicating that dark energy arose rapidly, such as from 
a phase transition.

\acknowledgments 

I thank colleagues at ICTP (Trieste), IFPU (Trieste), 
Roma La Sapienza (Rome), UiO (Oslo), LMU (Munich), IAA 
(Granada), US (Seville), UAM (Madrid), CGWU (Seoul), 
and KAS (Korea) for hospitality and discussions that 
helped motivate and crystallize this work. I am grateful 
to Kushal Lodha for computing the algebraic case posteriors.

\end{document}